\documentclass[twocolumn,showpacs,showkeys,preprintnumbers,amsmath,amssymb,prb,floatfix]{revtex4}
\usepackage{graphicx}
\usepackage{dcolumn}
\usepackage{bm}

\begin{document}


\title{Effect of in-plane magnetic field on the photoluminescence
spectrum of modulation-doped quantum wells and heterojunctions}

\author{B. M. Ashkinadze, E. Linder, E. Cohen}
\affiliation{Solid State Institute,Technion-Israel Institute of
Technology, Haifa 32000, Israel}

\email{borisa@tx.technion.ac.il}

\author {L. N. Pfeiffer}
\affiliation{Bell Laboratories, Lucent Technologies, Murray Hill, NJ 07974, USA}

\date{\today}

\begin{abstract}
The photoluminescence (PL) spectrum of modulation-doped
GaAs/AlGaAs  quantum wells (MDQW)  and heterojunctions (HJ) is
studied under a magnetic field ($B_{\|}$) applied parallel to the
two-dimensional electron gas (2DEG) layer. The effect of $B_{\|}$
strongly depends on the electron-hole separation ($d_{eh}$), and
we revealed remarkable $B_{\|}$-induced modifications of the PL
spectra in both types of heterostructures. A model considering the
direct optical transitions between the conduction and valence
subband that are shifted in k-space under $B_{\|}$, accounts
qualitatively for the observed spectral modifications. In the HJs,
the PL intensity of the bulk excitons is strongly reduced
relatively to that of the 2DEG with increasing $B_{\|}$. This
means that the distance between the photoholes and the 2DEG
decreases with increased $B_{\|}$, and that free holes are
responsible for the hole-2DEG PL.
\end{abstract}


\maketitle

\section{\label{sec:intro}  Introduction }

The low-temperature radiative recombination of the two-dimensional
electron gas (2DEG) with photoexited holes is an effective optical
probe of the many-body interactions and their modification under a
magnetic field that is applied perpendicularly ($B_{\bot}$) to the
2D-electron layer. Kinks in the $B_{\bot}$-dependence of the
photoluminescence (PL) peak energy,  the PL line broadening and
the intensity changes  for integral and fractional 2D-electron
filling factors $\nu=2\pi n_{2D} L_{B}$ ($n_{2D}$  is the 2DEG
density, $L_{B}$ is the magnetic length) were reported.
\cite{Gold,Heim,Pote}  The most remarkable PL-modifications were
observed in structure having a large 2D-electron - valence hole
(2D e-h) separation, $d_{eh}$, since then the holes weakly affect
the many-body interactions of the 2DEG. Examples are
asymmetrically modulation-doped, GaAs/AlGaAs quantum wells (MDQW)
(with a QW-width exceeding 20nm) \cite{Gold,Osb} and single
GaAs/AlGaAs heterojunctions (HJ).\cite{Tur,Dav,Nic}

In the latter case, the photoexcited electron-hole pairs are
rapidly separated by the built-in HJ electric field (over
distances of $d_{eh}>100$ nm within the entire undoped GaAs
layer). Due to the small 2D electron-hole wavefunction overlap,
their emission intensity is negligibly low and the PL spectrum of
high quality HJ's, in the absence or at low $B_{\bot}$ ($\nu>2$),
is dominated by emission of excitons from the undoped (p-type)
GaAs layer. \cite{Ash1,Nic}  However for filling factor $\nu<2$,
this excitonic PL transforms into a 2D e-h PL involving
transitions between the lowest e-h Landau levels. This changeover
in HJs was considered to reflect an increased 2D e-h wavefunction
overlap with increasing $B_{\bot}$ (at $\nu<$2)
\cite{Tur,Nic,Caj,Lee}, but its physical mechanism is not fully
understood. Recently, we proposed that the interaction of free
excitons in the GaAs buffer layer and the magnetized 2DEG forming
on the GaAs/AlGaAs interface, leads to an exciton dissociation
into 2D-electron and free hole at $\nu<$2. The excitons drift to
the 2DEG in the gradient of the built-in HJ electric field. Thus,
the exciton drift and its dissociation deliver free holes to the
2D-e. \cite{Ash2}

In order to elucidate the effect of the 2D-electron-hole
separation on the PL spectrum, we studied the PL of GaAs/AlGaAs
HJ's and MDQW's under a magnetic field that was applied parallel
to the 2DEG plane, $B_{\|}$. Extensive transport and
magnetoabsorption studies of the 2DEG under $B_{\|}$, were
reported, \cite{Koch,Eis,Sch,Tut}  but there are only a few
reports on the $B_{\|}$-effects on the 2DEG PL in MDQWs.
\cite{Whi,Ku,Hu}  A noticeable case is the effect of $B_{\|}$ on
the spatially indirect exciton PL in biased double quantum wells.
\cite{but}  We report on drastic $B_{\|}$-induced PL spectral
changes in high quality 25nm-width MDQW's and in HJ's as well as
on their dependence on the 2D e-h separation. We present a model
that is based on the conduction and valence subband realignment
under $B_{\|}$ that accounts qualitativily for the observed
spectral modifications. The effect of a $d_{eh}$ decrease with
$B_{\|}$  on the PL spectrum of  HJs, is also considered.

\section {Model. PL spectral modifications induced by $B_{\|}$ }

An in-plane $B_{\|}$ that is applied along the x-axis, creates a
crossed fields configuration with the perpendicular, built-in
electric field $E_{\bot}$ (directed along the z-axis) that exists
in the asymmetrically modulation-doped structures containing a
2DEG. This causes an in-plane electron (hole) drift (in the
y-direction, perpendicular to $B_{\|}$), resulting in a
deformation of the subband energy surfaces, $\epsilon_{k}$.
\cite{St} In particular, the conduction subband minimum shifts to
a higher wave vector, $k_{y}^{B}$=$d_{eh}/L_{B}^{2}$ (where
$L_{B}$ is the magnetic length) and the in-plane electron
effective mass ($m_{ey}$) increases along the y-direction.
\cite{St,m}  Thus, an indirect bandgap appears, and the 2DEG-free
hole PL spectrum that originates in the direct optical
transitions, is strongly modified.

In order to describe the spectral modifications, we use the
general expression for the spectrum of the 2De-h radiative
recombination. \cite{Bast} The PL intensity at a photon energy
$\hbar\omega=E_{g}+\varepsilon_{e} + \varepsilon_{h}$  is
\begin{eqnarray}  \label{1}
I(\hbar\omega)=\int\int{dk_{x}dk_{y}f(\varepsilon_{h})f(\varepsilon_{h})
\delta(\hbar\omega-E_{g}-\varepsilon_{e}
- \varepsilon_{h})} \
\end{eqnarray}
Here $E_{g}, \varepsilon_{e}, \varepsilon_{h}$   are the band gap,
electron and hole in-plane energies, and $f(\varepsilon_{e}),
f(\varepsilon_{h})$ are the Fermi distribution function for
2D-electrons and the Boltzmann distribution function  for
nondegenerate holes, respectively. In the direct band gap limit
($B_{\|}$ =0), all optical transitions with a given $\hbar\omega$
 occur at $\hbar k_{e}=\hbar k_{h}=(2\mu(\hbar\omega-E_{g}))^{1/2}$ ($\mu$ is the
reduced electron-hole effective mass). Thus, the PL spectrum is
described  by $I(\hbar\omega)$$\propto$$ f(\varepsilon_{e})
f(\varepsilon_{h})$.

In the presence of $B_{\|}$, direct optical transitions occur
between the valence and conduction subbands that are displaced in
$k$-space away from each other by $k_{y}^{B}=d_{eh}eB_{\|}$/
$\hbar c$. Thus,  the momentum and energy conservation laws
require that for given $\hbar\omega$ and $k_{y}$:
\begin{eqnarray}\label{2}
k_{x}=[\frac{2\mu}{%
\hbar^{2}}(\hbar\omega-E_{g}-\frac{\hbar^{2}k_{y}^{2}}{%
2m_{h}}-\frac{\hbar^{2}(k_{y}-k_{y}^{B})^{2}}{%
2m_{ey}})]^{1/2}.
\end{eqnarray}
and conduction (and valence) band states of $\emph{different}$
energies participate in the optical transitions at the same photon
energy. The optical transitions at $\hbar\omega$ involve states
with $k_{y}$ varying between $k_{y1}$ and $k_{y2}$  that are the
roots of the Eq.2 (at $k_{x}=0$). Then, integrating  Eq.1 once, we
obtain
\begin{eqnarray}\label{3}
I(\hbar\omega)= \int_{k_{y1}}^{k_{y2}}dk_{y}
\frac{f(\varepsilon_{h})f(\varepsilon_{h})\delta(\hbar\omega-E_{g}-\varepsilon_{e}
-\varepsilon_{h})}{%
k_{x}(k_{y})}
\end{eqnarray}
where  $f(\varepsilon_{h})=$exp$(-\varepsilon_{h}/k_{B}T_{h})$ ,
$f(\varepsilon_{e})=
(1+$exp$((\varepsilon_{e}-E_{F})/k_{B}T_{e})^{-1}$. $T_{e}$ and
$T_{h}$ are the effective electron and hole temperatures, and
$E_{F}= E_{F}^{0}m_{ex}/(m_{ex}m_{ey})^{1/2}$ is the Fermi energy
in the presence of $B_{\|}$. We note that $E_{F}<E_{F}^{0}$  since
$m_{ey}$ increases with $B_{\|}$. \cite{Ku,Sch,Tut}

In Figure~1, numerically calculated PL spectra are presented for
several $k_{y}^{B}$ values that correspond to increasing  $B_{\|}$
values. The PL peak intensities are obtained from the condition of
B-independent spectrally-integrated PL. In the presence of
$B_{\|}$, the lowest PL energy shifts by
\begin{eqnarray}\label{4}
\epsilon_{B}= \frac{(\hbar
k_{y}^{B})^{2}} {%
2(m_{ey}+m_{h})}= \frac {e^{2}d_{eh}^{2}B^{2}_{\|}
}{%
2c^2(m_{ey}+m_{h})}
\end{eqnarray}
and the PL spectrum is strongly deformed, particularly at large
$k_{y}^{B}$. This results from a change of number of the occupied
free-hole states participating in the recombination process. For
example, as $k_{y}^{B}$ increases, the direct optical transitions
between the 2D-electrons at $E_{F}$ and the lowest energy, highly
populated valence hole states, have become available. This leads
to a pronounced PL intensity enhancement at $E_{F}$, as  can be
seen in Fig. 1. The PL spectral evolution with increasing
$k_{y}^{B}\propto B_{\|}$   is shown in Figs.~1a, b for two values
of the 2DEG density. These spectra demonstrate that the main
effect of $B_{\|}$ is not an enhanced diamagnetic shift (see Eq.3)
as was considered before, \cite{Whi} but the drastic modification
of the entire 2De-h PL spectrum. For example, the lowest optical
transition shifts by $\epsilon_{B}\simeq 2$meV at
$k_{y}^{B}\simeq1.8\cdot10^{6}$cm$^{-1}$ ($B_{\|}$ =7T and
$d_{eh}$=18 nm), while the PL peak-energy shift depends on the
2DEG density and reaches ~7meV at $n_{2D}=2.10^{11}$ cm$^{-2}$
(see Figs.~1a, b).

The energy distribution of the photoexcited free holes
participating in the 2De-h PL, strongly affects the PL spectrum
under $B_{\|}$. The energy distribution of the holes can be
different from that corresponding to the lattice temperature
($T_{L}$) because the radiative recombination rate is higher than
the energy relaxation rate in the MDQW at low temperatures.
\cite{Heim} In order to demonstrate the effect of the
nonthermalized holes on the PL, we display the PL spectra
calculated for the effective hole temperature $T_{h}$ = 4K (dashed
lines in Fig.~1). $T_{e}$ is taken to be equal to $T_{L}$=1.9K,
since the photoelectron rapidly loses its energy by the
electron-electron scattering process occurring in the dense 2DEG.
Due to the heavier hole mass and the spatial separation of the
holes and 2DEG, the efficiency of the hole-2DEG energy relaxation
is lower, and $T_{h}$ is taken to be different of $T_{e}$. One can
see in Fig.~1 that the high-energy valence states occupied by
nonthermalized holes, result in a pronounce PL spectral
modification under increased $B_{\|}$.

The PL spectra shown in Fig.~1 were calculated with $m_{ey}=
m_{ex}$. The dotted curve in Fig.~1b shows the calculated PL
spectrum for  $m_{ey} = m_{ex} +b k_{y}^{B}$ -dependence with
$k_{y}^{B}=1.4\cdot10^{6}$cm$^{-1}$ ($b=10^{-7}$ is a numerical
coefficient). The larger $m_{ey}$ leads to a PL band narrowing and
a low-energy shift of the PL peak due to Fermi energy decrease.
Thus, the strong $B_{\|}$ effect on the 2De-h PL spectrum is
predicted by this simple model. For a MDQW in which $n_{2D}$ can
be varied, larger spectral modifications are expected at higher
$n_{2D}$ since $d_{eh}$ increases with $n_{2D}$ due to the
increased built-in electric field. Our analysis does not include
the "usual" diamagnetic shift. The value of this small shift ($<1$
meV at 7T) is close to the exciton diamagnetic shift under
$B_{\|}$ as measured for undoped 20nm wide QW (see below, Fig.~3).

The effect of $B_{\|}$ is expected to be different in wide HJ's,
since $d_{eh}$ is large, and it varies with increasing $B_{\|}$.
The simplest estimate of $d_{eh}$ in a HJ at $B_{\|}$  =0, can be
obtained by  $d_{eh}\simeq V_{h}\tau =\mu_{h}E_{\bot}\tau$ where
$V_{h}$ is the valence hole drift velocity in the HJ electric
field  $E_{\bot}$ ( $\mu_{h}$ is the hole mobility) and $\tau$ is
the characteristic recombination time of the hole (due to capture
by charged acceptors in the buffer p-type  GaAs layer). Taking
$\mu_{h}=10^4$ cm$^2$/Vsec  \cite{Ad} and a minimal value of
$E_{\bot}$ =$10^3$ V/cm and $\tau=10^{-10}$s, we obtain $d_{eh}>
10^{-4}$cm. The photoexcited holes are thus accumulating at a
large distance where $E_{\bot}$ diminishes. Therefore, $d_{eh}$ is
of the order of the GaAs buffer layer width ($1\mu$), and it is
much larger than that in the MDQWs.

In the presence of $B_{\|}$, the hole drift from the interface is
slowed down, since it exhibits a helical motion along the
y-direction (in $E_{\bot}$- $B_{\|}$ crossed fields
configuration). The hole drift velocity can be written (in a
classical approach) as: $V^{B}= \mu_{h}E_{\bot}/[1+(\mu_{h}
B_{\|}/c)^{2})]$, ($c$ is the light velocity).  Then, for
$B_{\|}>$1T ($\mu_{h}B_{\|}/c\simeq1$), $d_{eh}$ strongly
decreases, reaching values $<10^{-5}$cm at $B_{\|}>$ 2T. Thus, the
spatial distribution of photoexcited holes (in the buffer GaAs
layer) is mainly determined by the incident light penetration
depth. The hole density near the 2DEG layer increases while the
density of holes situated away from HJ decreases. Fewer holes are
available to form excitons in the buffer GaAs layer, and the
exciton PL intensity decreases while that of the 2De-h PL is
enhanced with increasing $B_{\|}$ . It is important to underline
that the discussed PL modifications in $B_{\|}$ are only relevant
for free holes that recombine with momentum conservation. In the
case of recombination of localized holes with 2DEG, the spectral
PL modifications are expected to be small since indirect optical
transitions without $k$-conservation are allowed.

\section{Experimental results and discussion}

The PL spectroscopic study was performed on several GaAs/AlGaAs HJ
and MDQW  samples grown by molecular beam epitaxy. The HJ samples
have a thick GaAs buffer layer (widths of $1\mu$ ) and the MDQW
samples have a single 25nm-wide QW.  The 2DEG densities and dc
mobilities at 4K vary in the ranges of  $n_{2D} = (0.7-3)\cdot
10^{11}$ cm$^{-2}$ and $\mu =(1 - 4)\cdot10^{6}$ cm$^{2}$/Vsec,
respectively. Photoexcitation was done by illumination with a
Ti-sapphire laser light (photon energy of 1.56 eV) or by a He-Ne
laser. The incident light intensity was kept below
$10^{-2}$W/cm$^{2}$. The He-Ne laser photon energy (1.96eV) is
greater than the band gap of the AlGaAs barrier, thus, $n_{2D}$
can be reduced due to optical depletion by increasing the He-Ne
laser intensity. \cite{Wor,Ash3} The PL spectra were measured with
a high resolution by using a double spectrometer equipped with a
CCD camera. The samples were immersed in liquid He at temperature
$T_{L}$= 1.9K, and photoexcitation and PL detection were performed
perpendicularly to the 2DEG plane under an in-plane $B_{\|}$.

 Figs.~2a-c display the PL spectral evolution with increasing $B_{\|}$
measured on two MDQW samples. At $B=0$, PL spectra can be well
described by a simple product of the distribution functions for
the 2D electrons of density $n_{2D}$  at the electron temperature
$T_{e}$=$T_{L}$ and nondegenerate holes with effective $T_{h}$
(Eq.1). \cite{Ash4}  The 2DEG Fermi energy is then estimated from
the 2De-h PL bandwidth. Upon applying $B_{\|}$, the PL spectra
show remarkable modifications: a high-energy shift, intensity
redistribution, and band narrowing. Fig.~2a presents the PL
spectra of the MDQW with $n_{2D}^{0}=3\cdot10^{11}$cm$^{-2}$ at
$B_{\|}$=0, 4.5 and 7T and the calculated spectra for
$k_{y}^{B}=0,  1,   1.4$ and $1.6\cdot10^{6}$cm$^{-1}$ for
$T_{e}$=1.9K, $T_{h}$= 4K. A $d_{eh}$ value of 15 nm can be
estimated from the comparison of these spectra. Figs.~ 2b, c show
the effect of varying $n_{2D}$ on the PL spectral evolution under
$B_{\|}$. The PL spectra are measured on the same MDQW with
$n_{2D}^{0}=1.8\cdot10^{11}$cm$^{-2}$ under two He-Ne laser
intensities (under optical depletion). As $n_{2D}$ is reduced,
less spectral modifications are observed since $d_{eh}$ decreases.
The total width of the 2De-h PL spectra ($ \propto E_{F}$),
decreases with $B_{\|}$. At $B_{\|}$=7T, $E_{F}$ decreases by
1.5-1.3 times as a result of an electron mass
enhancement.\cite{Sch,Tut}

The observed PL spectra display the main features predicted by the
simple model of Sec II.  This model does not accounts for the
$B_{\|}$-effect on the electron(hole) wavefunctions, but we assume
that the discrepancy between the calculated and observed PL
spectra  may result from the nonthermalized hole distribution
function, which is not described by the Boltzman distribution with
the effective $T_{h}$.

The $B_{\|}$ dependencies of the integrated PL intensity ($J$) and
the energy $E_{B}$ that was obtained by extrapolating the PL
intensity of low energy part of spectrum to zero, are shown in
Fig.~3. We note that the integrated PL intensity varies only
slightly with $B_{\|}$.  $E_{B}$ is presented  for a MDQW at two
2DEG densities, $n_{2D}^{0} = 1.8\cdot 10^{11}$ cm$^{-2}$ and
$n_{2D} \approx 1 \cdot 10^{11}$ cm$^{-2}$  (curves 1 and 2,
respectively). These dependencies are compared with the behavior
of the exciton PL peak energy versus $B_{\|}$ that was measured in
the 20nm-wide undoped QW (curve 3). $E_{B}\simeq\epsilon_{B}$ in
the MDQW varies with $n_{2D}$, and it strongly depends on $B_{\|}$
when compared with the exciton diamagnetic shift in undoped QW.

Fig.~4 displays the PL spectra of  the HJ sample upon applying a
\emph{perpendicular} magnetic field, $B_{\bot}$ (Fig.~4a) and a
parallel field, $B_{\|}$ (Figs.~4b, c). Figs.~4b and c show the PL
spectra for the same HJ ($n_{2D}^{0}=1.6\cdot 10^{11}$ cm$^{-2}$)
under photoexcitation with $E_{L}$=1.52  eV  and 1.96 eV (under
optical depletion), respectively. At $B$=0, the PL spectrum
consists of two strong narrow lines originating in free and bound
excitons of GaAs buffer layer. \cite{Ash1}  With decreasing
$n_{2D}$ (Fig~4c), the built-in electric field decreases, holes
are situated closer to the heterointerface, and the broad 2De-h PL
band appears at $B=0$. Under photoexcitation that does not vary
$n_{2D}$ (Figs.~4a, b), the 2De-h PL can be detected as a
low-energy low-intense background PL. These broad PL bands are due
to a radiative recombination of the photoexcited holes and
2D-electrons whose wavefunctions weakly overlap at $B=0$. At
$B_{\bot}\simeq$ 3.2 T ($\nu\sim$2), the PL changeover occurs, and
the 2De-h PL intensity sharply enhances while the exciton PL
intensity decreases (Fig.~4a). \cite{Ash2} Another drastic 2De-h
PL modification is observed near $\nu\sim$1 ($B_{\bot}\sim$
6.5-7T). \cite{Osb,Ash2}

The PL evolution with applying in-plane $B_{\|}$ is shown in
Figs.~4b, c. Starting from low $B_{\|}$, an increase of the 2De-h
PL (low-energy PL tail) intensity  and  an narrowing of the band
with $B_{\|}$ are clearly observed. A redistribution between the
2De-h and exciton integrated PL intensities with increasing
$B_{\|}$ is presented in inset of Fig.~5.  A \emph{smooth
}changeover from the exciton to the 2D-e-h PL under
$B_{\|}\simeq1T$ is revealed.  This is in contrast to the case of
the sharp, 2DEG density-dependent changeover observed under
$B_{\bot}$.\cite{Ash2}

The 2D-e-h PL intensity enhancement seen in Figs.~4b, c, results
from the $B_{\|}$-induced delivering of free holes to the
heterointerface while the narrowing and spectral PL shift is
caused by the effect of $B_{\|}$ on the PL spectral shape. Indeed,
with increasing $B_{\|}$, the photoexcited holes are swept away
for smaller distances, and increased number of the holes can
recombine with 2D-electrons giving rise to the 2De-h PL. There is
a set of the $d_{eh}$ -values because the holes are spatially
distributed in GaAs buffer layer. This leads to a specific PL
spectrum having a long low-energy tail since the PL intensity
strongly reduces with increasing $d_{eh}$.  As  $B_{\|}$
increases, $d_{eh}$ decreases, however $d_{eh}$ is large enough so
that $k_{y}^{B}$ is high, and the PL line shape at $B_{\|}$=7T
corresponds to the PL spectrum calculated for
$k_{y}^{B}>2.10^{6}$cm$^{-1}$ as one can see in Fig. 1.

In Fig.~5, we compare the HJ PL spectra under parallel (curves
1,4), normal (curve 2) and $45^{0}$-tilted magnetic field (curve
3). The spectra presented by curves 1 and 2 are obtained at
$B_{\|}$= 5T and $B_{\bot}$=5T, respectively. The high-energy,
excitonic part of both spectra are at the same energies, and it is
independent of the B-orientation. The excitonic part of the
spectrum  under tilted magnetic field ($B_{\|}=B_{\bot}$=5T, curve
3)is similar to that measured at $B_{\|}$=7T (curve 4). Both these
facts give evidence to the bulk nature of the high-energy part of
the PL spectrum in the HJ. The 2De-h parts of the PL spectra
obtained under $B_{\|}$ and $B_{\bot}$ (curves 1 and 2) are
completely different, and these differ from that observed under
tilted magnetic field. In the latter case, the 2De-h PL originates
in the radiative recombination from the lowest e-h Landau levels
(due to $B_{\bot}$-component) while an additional spectral shift
$\varepsilon_{B}$ is caused by $B_{\|}$-component. The value of
this shift is of 1.8meV, and the 2De-h separation of 25 nm  can be
estimated by using Eq. 3. Thus applying $B_{\|}$=5T, leads to a
reduced hole drift that results in the strong 2De-h PL
enhancement.

\section{Conclusions}

In conclusion, our study shows that in-plane magnetic field leads
to a remarkable spectral modification of the 2DEG-hole PL in MDQWs
and SHJs. The smooth intensity redistribution between the PL of
the bulk excitons and of the 2DEG is revealed in HJs. This is
caused by the  effect of $B_{\|}$ on the free hole distribution in
the HJ. The 2De-h PL evolution studied upon applying $B_{\|}$,
evidences that free holes are responsible for this emission in the
high quality HJs. Thus, we can conclude that the sharp PL
changeover observed at $\nu\simeq$2 in HJs under perpendicularly
applied $B_{\bot}$, is induced by a threshold spatial
redistribution of the free holes whose possible physical mechanism
was recently proposed. \cite{Ash2}

\begin{acknowledgments}
We thank V. Yudson for the fruitful discussions. The research at
the Technion was done in the Barbara and Norman Seiden Center for
Advanced Optoelectronics. B. M. A. acknowledges the support of a
grant under the framework of the KAMEA Program.
\end{acknowledgments}


\newpage

\begin{figure}
\caption {The calculated PL spectra for two $n_{2D}$ values at
increasing $k_{y}^{B}=d_{eh}eB_{\|}/ \hbar c$. $m_{ex}=0.067
m_{0}$, $m_{ey}=[0.067(1+10^{-7}k_{y}^{B})]m_{0}$,
$m_{h}=0.35m_{0}$, $T_{L}=T_{e}=$1.9 K. Solid (dashed) lines are
for $T_{h}$=1.9K ($T_{h}$=4K), respectively. Dotted curve in
Fig.1b shows the effect of increased $m_{ey}$. $k_{y}^{B}$-values
are shown near the curves. }
\end{figure}
\begin{figure}
\caption  {2DEG PL spectral evolution under a parallel magnetic
field, $B_{\|}$=0, 4.5, 7T. a). 25-nm wide MDQW with
$n_{2D}^{0}\simeq3\cdot10^{11}$cm$^{-2}$ (solid lines). The fitted
spectra with correponding $k_{y}^{B}$-values are presented by
dotted lines. b, c) 25-nm wide MDQW with
$n_{2D}^{0}\simeq1.8\cdot10^{11}$cm$^{-2}$ at two He-Ne laser
light intensities, $I_{L}$: $I_{L1}$(b) is less than $I_{L2}$
(c).}
\end{figure}
\begin{figure}
\caption {$B_{\|}$-dependencies of the $E_{B}$-shift for the MDQW
at two $n_{2D}$: curve 1 is obtained at higher $n_{2D}$ than curve
2. Curve 3 - exciton PL peak energy shift for an undoped QW. Curve
4 - integrated PL intensity versus $B_{\|}$ for the MDQW.}
\end{figure}

\begin{figure}
\caption {The PL spectral evolution in the HJ upon applying
$B_{\bot}$ (a) and $B_{\|}$ (b, c). a and b are measured under
photoexcitatiion at $E_{L}$=1.52 eV , (c) - at 1.96 eV }
\end{figure}
\begin{figure}
\caption { PL spectra of HJ. 1 and 4 - under $B_{\|}$ =5 and 7T,
respectively, 2- under $B_{\bot}$=5T  and 3 - under a
$45^{0}$-tilted magnetic field ($B=7$T, $B_{\|}=B_{\bot}$=5T).
Inset: Integrated intensity of the exciton and 2De-h PL versus
$B_{\|}$. }
\end{figure}

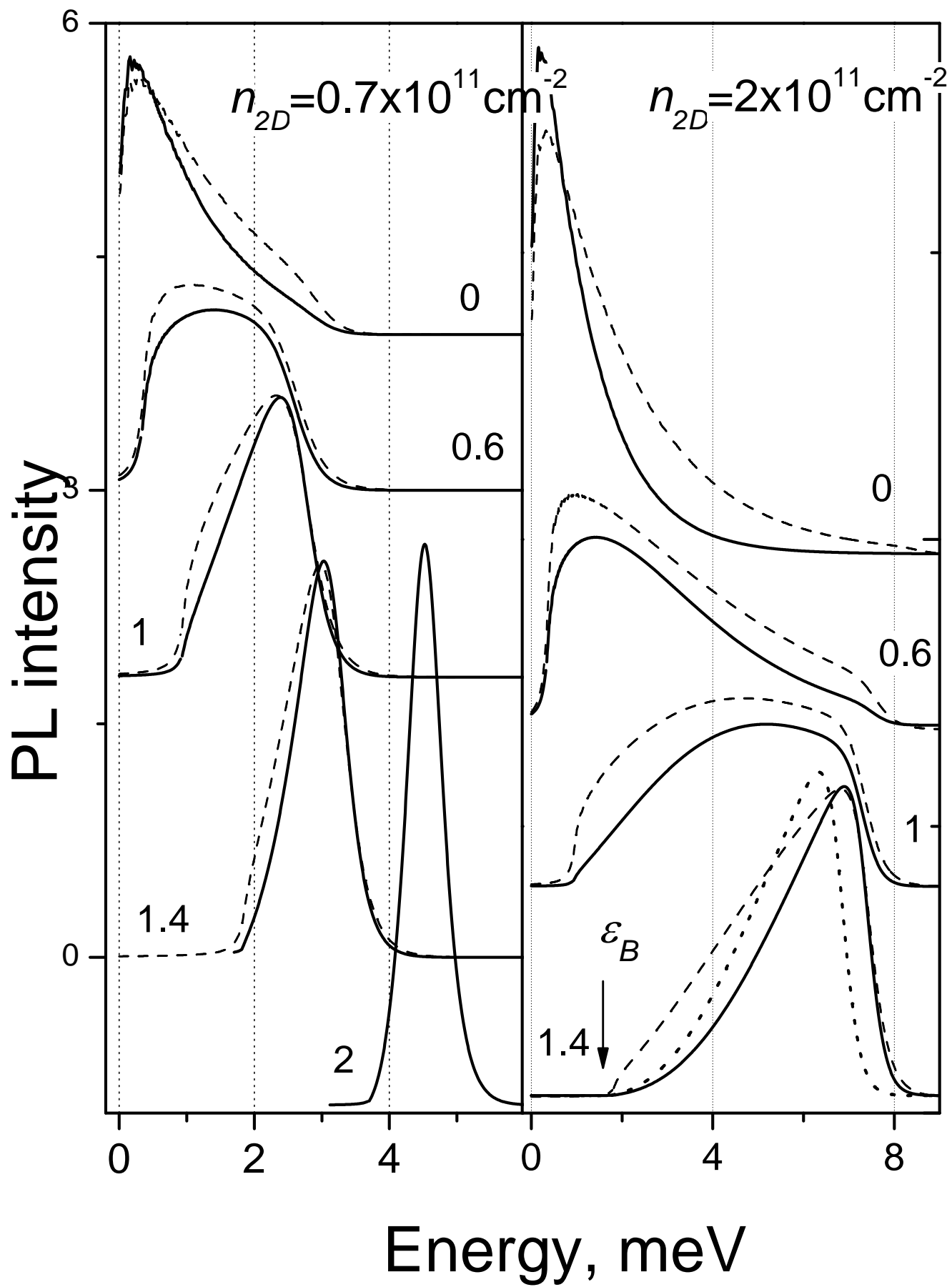
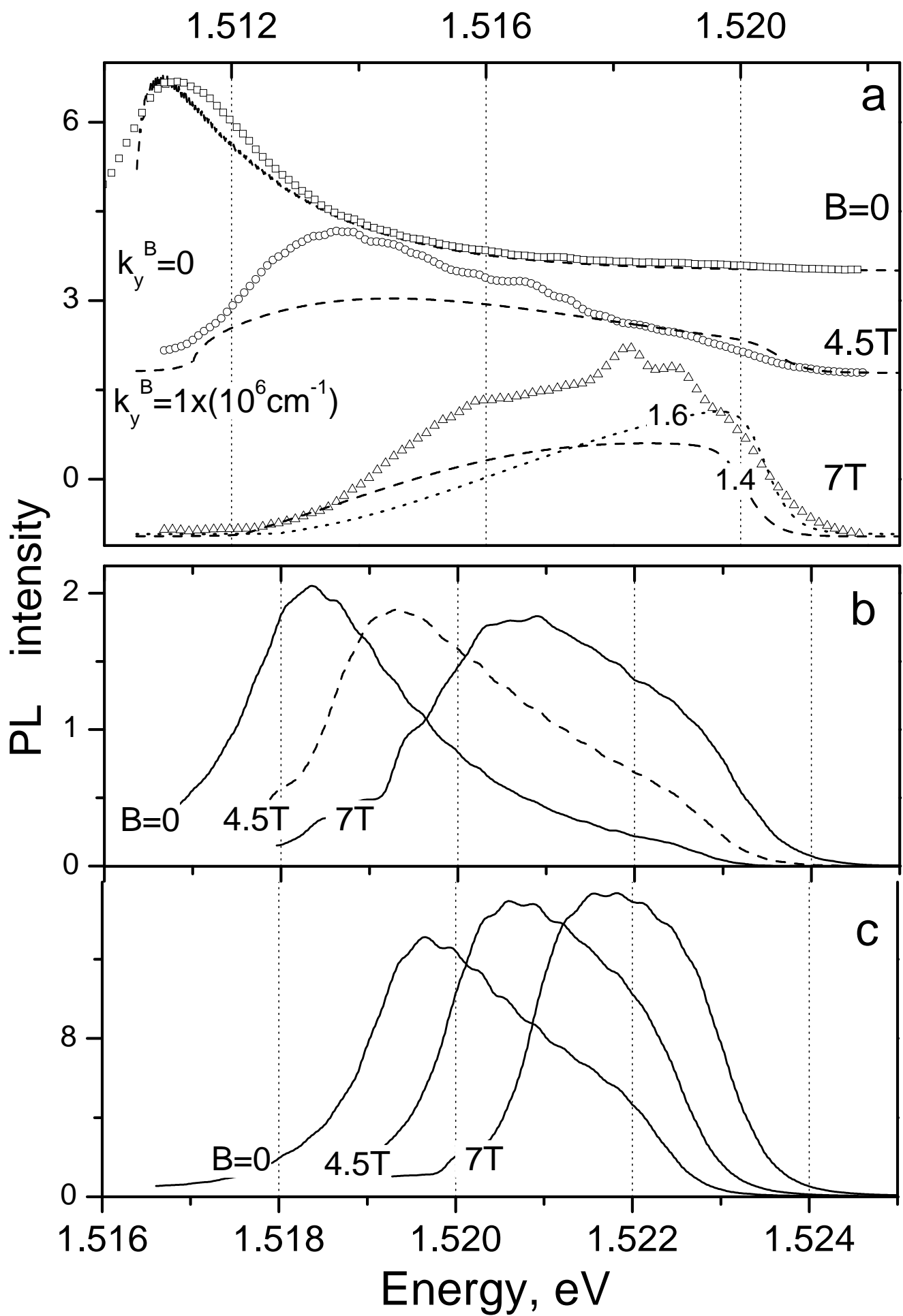
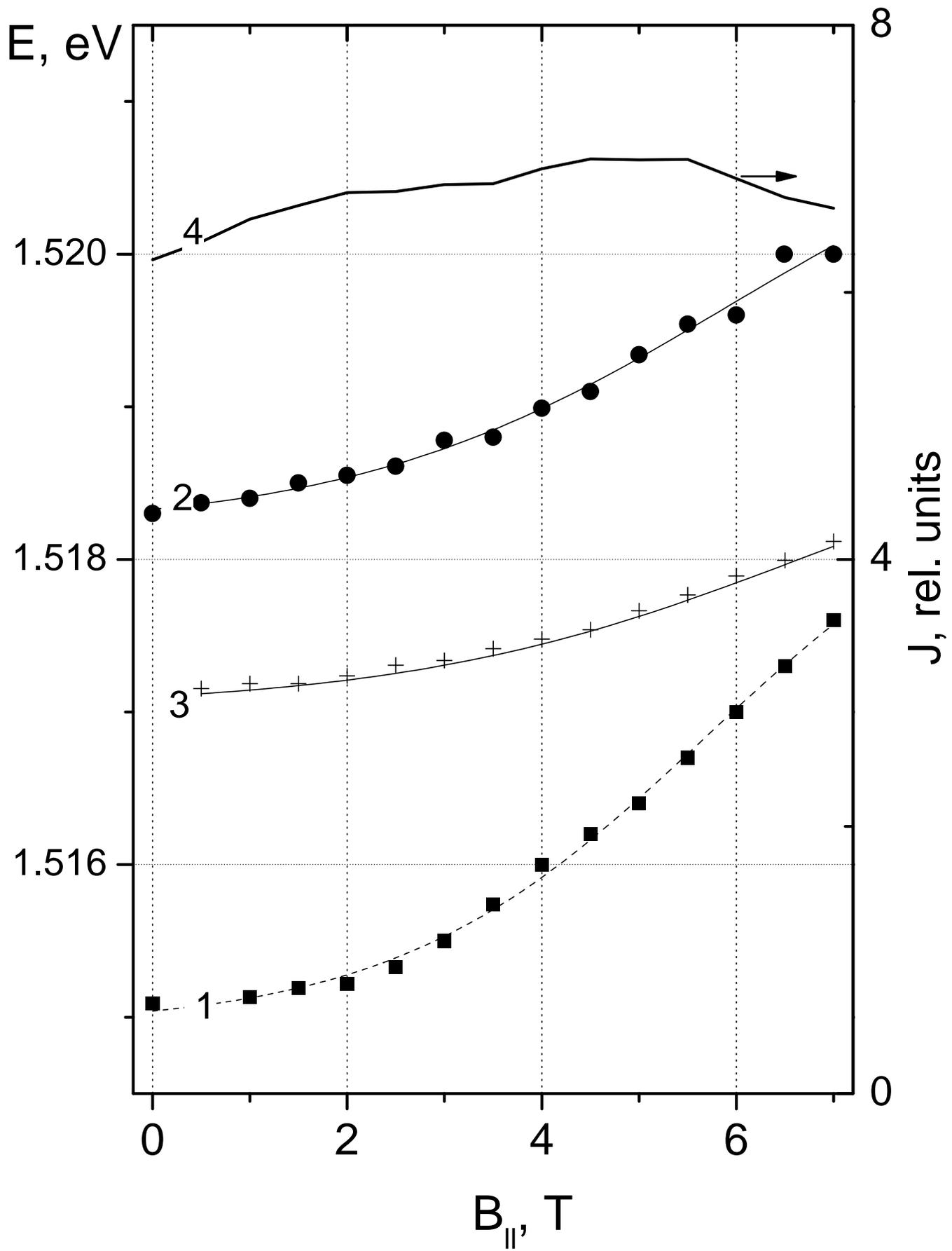
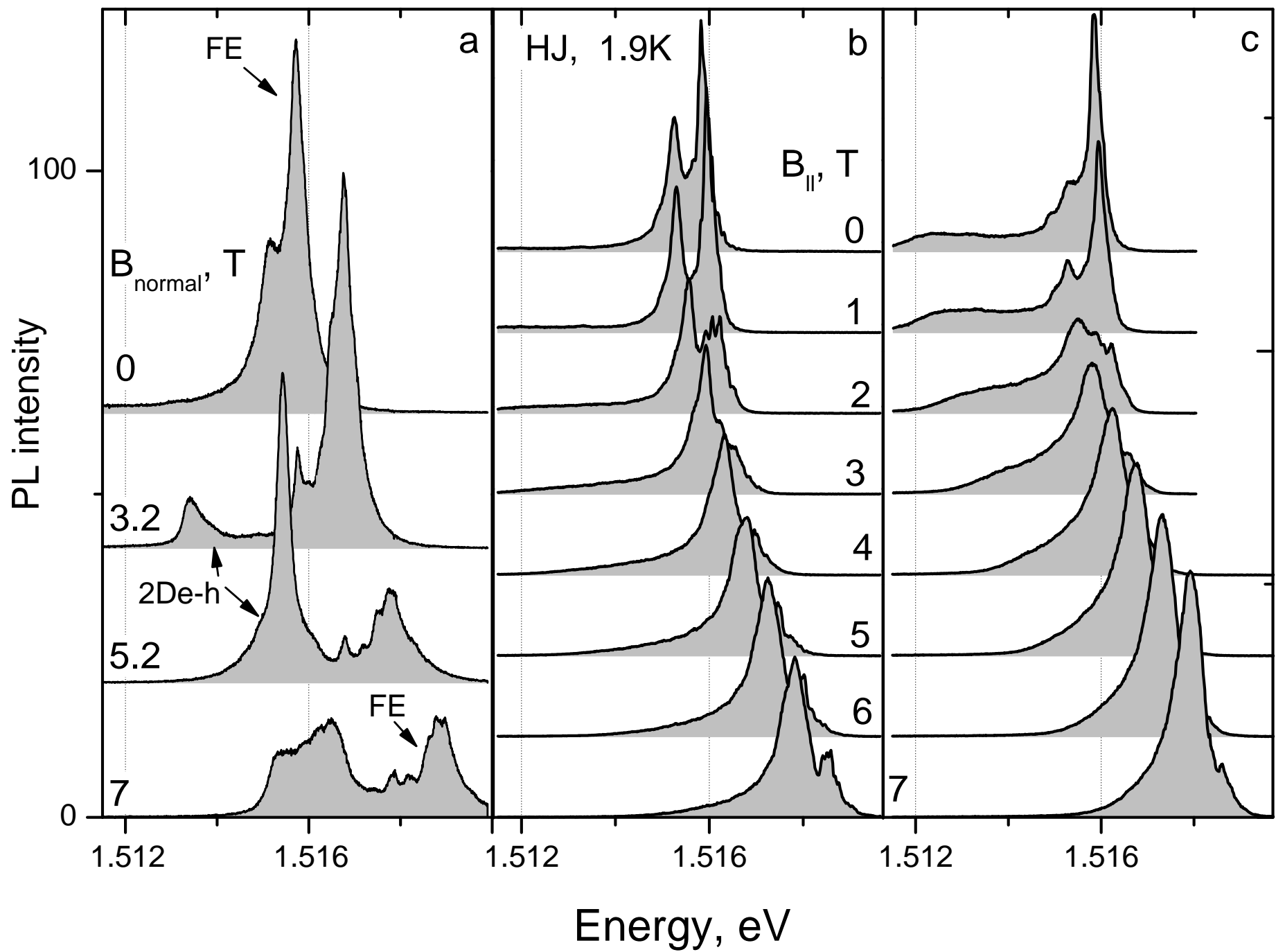
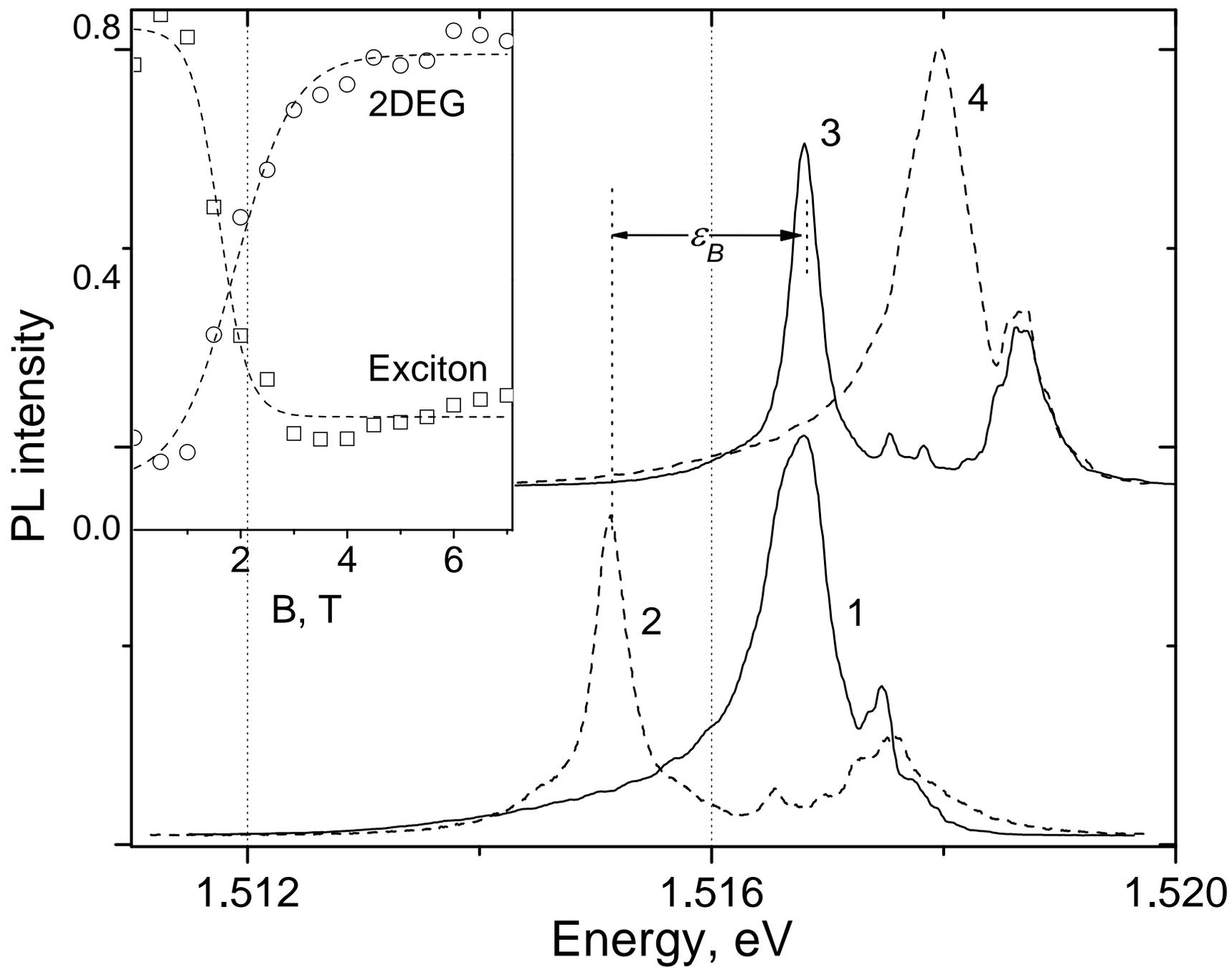
\end{document}